\documentclass[reprint,amsmath,amssymb,aps, prl]{revtex4-2}

\usepackage{graphicx}
\usepackage{dcolumn}
\usepackage{bm}
\usepackage[final]{changes}
\usepackage{color}

\begin{document}

\preprint{APS/123-QED}

\title{Topological Dissipation as the Missing Link in Multiscale Polymer Dynamics}

\author{Xu-Ze Zhang}
 \altaffiliation{These authors contributed equally to this work}
\author{Rui Shi}
 \altaffiliation{These authors contributed equally to this work}
\author{Ming-Ji Fang}
\author{Zhong-Yuan Lu} 
\author{Hu-Jun Qian}
 \email{hjqian@jlu.edu.cn}
\affiliation{
 State Key Laboratory of Supramolecular Structure and Materials, College of Chemistry, Jilin University, Changchun 130021, China.
}%

\date{\today}

\begin{abstract}

We identify \emph{topological dissipation}---momentum transport along the polymer backbone with Ising-type exponential decay ($\sim \exp(-\Delta n/n_\text{d})$)---as the missing link connecting atomistic and mesoscale dynamics. Simulations of four polymers reveal that dynamical correlation length $n_\text{d} \approx n_\text{k}/3$ (Kuhn length $n_\text{k}$), enabling a coarse-grained framework that \emph{explicitly separates} topological (intrachain) and spatial (interchain)  dissipation channels without temporal memory kernels. The approach quantitatively reproduces dynamics from segmental relaxation to chain diffusion, solving the long-standing memory preservation challenge in Markovian coarse-graining. Our results establish topology-mediated dissipation as a key mechanism for polymer dynamics.

\end{abstract}

\maketitle
The chain-like topology of polymers constitutes the fundamental origin of their dynamical complexity. In both classical theories and molecular simulations, polymers are commonly simplified as coarse-grained (CG) bead-spring models to capture the chain-like characteristic \cite{de1979scaling, doi1988theory, grest1986molecular}. As a paradigmatic example of the Rouse model and its extensions\cite{rouse1953theory, watanabe1999viscoelasticity, likhtman2005single, diddens2015chain}, harmonically connected beads exhibit liquid-like dynamics, with each monomer experiencing local friction and independent Brownian noise---characterized by either uniform or position-dependent friction coefficients.
\par
However, such a simplified model overlooks a key molecular-scale dynamical effect: the momentum of each monomer (composed of multiple atoms with internal degrees of freedom) becomes correlated through covalent bonds along the polymer backbone, enabling collective momentum propagation through the chain. In atomistic representations, each atomic degree of freedom (DoF) contributes non-negligibly to the momentum of individual monomers. Crucially, these dynamical correlations between atomic DoFs of connected monomers are systematically eliminated during the coarse-graining procedure. We term such correlation arising from atomistic motions along the chain backbone as \emph{topological dissipation}. This aligns with our findings in polymer nanocomposite~\cite{chentaoNC2019} where localized accelerations near nanoparticles propagate with Ising-like correlation decay along the chain backbone, reducing viscosity in a chain-length-dependent manner, confirming that chains act as dissipative waveguides rather than passive elastic strings.

\par
From the perspective of CG molecular dynamics simulations, the Mori-Zwanzig (MZ) formalism~\cite{mori1965continued,zwanzig1961memory} provides a rigorous framework to derive exact equations of motion (EoM) for CG variables, it encodes topological dissipation effects into opaque space-time dependent memory kernels, obscuring the explicit influence of topological dissipation on dynamics. Modern parameterization strategies—including machine learning~\cite{huan2023construction,ge2024data}—have numerically improved kernel reconstruction for various systems~\cite{rudzinski2019recent, klippenstein2021introducing, schilling2022coarse, jung2017iterative, jung2018generalized, jung2023dynamic, li2015incorporation, li2017computing, klippenstein2021cross, klippenstein2023bottom, han2018mesoscopic, han2021constructing}. However, it still remains a major challenge to extract and apply the memory kernel in polymer systems. For instance, the memory effect might be as long as the relaxation time of the chain. Therefore, one would expect a strong time dependence of $\tau\sim N^\nu$ for the memory kernel, with the exponent $\nu$ can be as large as $\sim$3.4. As we demonstrated in our previous general Langevin equation (GLE) model \cite{zhang2024chemically}, although using a reasonable cutoff in time for the memory kernel can effectively capture high-frequency motions, there are still some deviations in collective dynamical properties at larger time scales, i.e., intermediate scatter function, van Hove functions etc.

\par
In this work, we begin with atomistic molecular dynamics simulations of several distinct polymer systems, through which we first establish the universal existence of topological dissipation---the directed propagation of momentum fluctuations along polymer backbones.  Building upon these atomic-scale insights, we develop a CG framework that explicitly incorporates this physics through two coupled dissipation channels: intramolecular dissipation, modeled via a 1D Ising-type formulation to capture such topological exponential decay mechanism, and intermolecular dissipation described through conventional Langevin friction. We take polypropylene melt as an example, this approach successfully reproduces the full spectrum of dynamical behavior in CG model, from segmental relaxation to chain diffusion, without invoking non-Markovian memory kernels. The remarkable agreement between CG predictions and reference atomistic models demonstrates that chain connectivity intrinsically encodes the essential dissipation physics underlying multiscale polymer dynamics.
\par
\par

\begin{figure}[h]
	\includegraphics[width=0.9\linewidth]{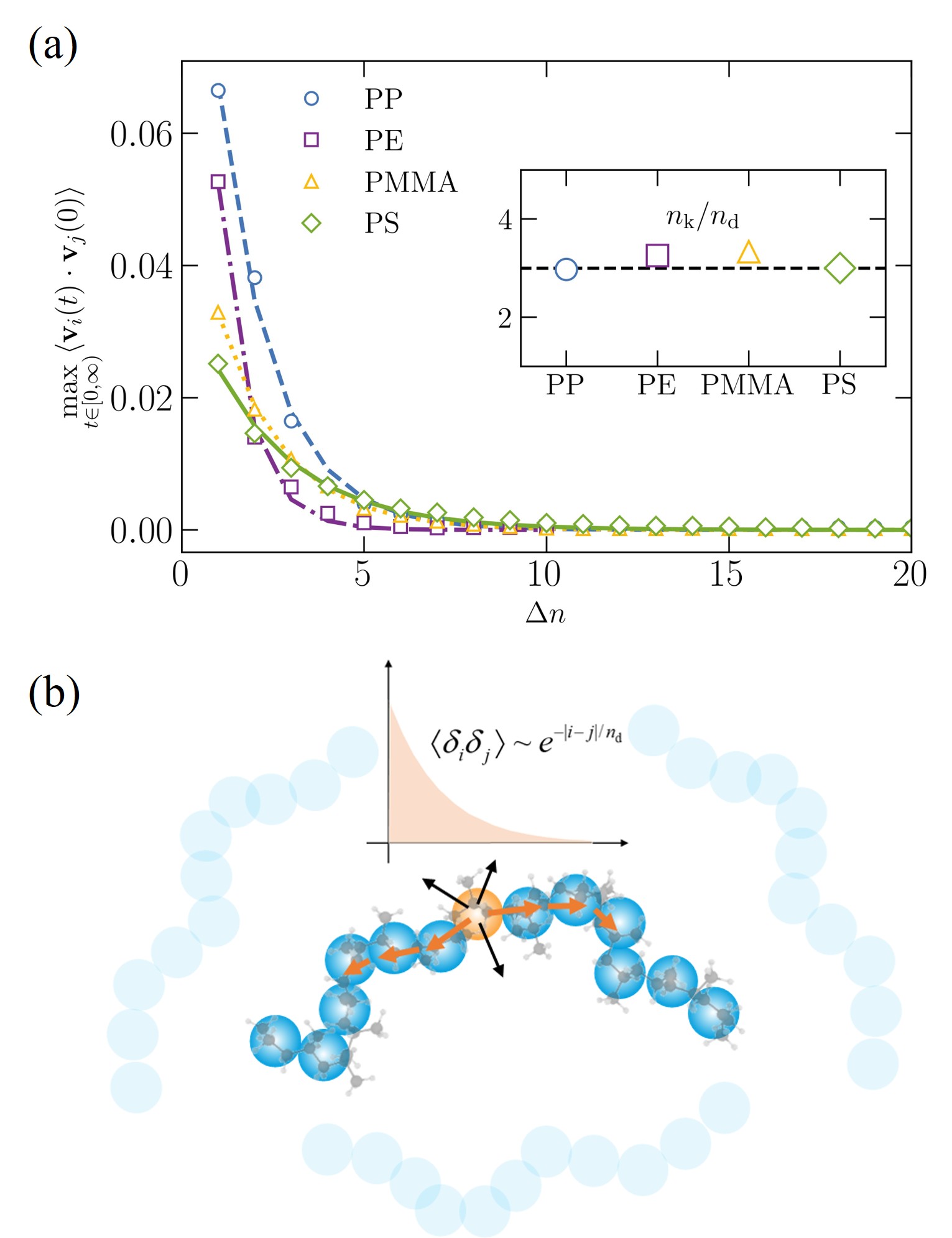}
	\caption{(a) The maximum value of the VCCF, between two CG beads separated by $\Delta n$ beads along a polymer chain, is calculated from atomistic simulations for four different polymer melts. The various lines depict exponential fitting with $B\exp{(-\Delta n/ n_\text{d})}$. (b) Schematic illustration of Eq.~\ref{eq:EOM}. Orange arrows denote topological dissipation, while black arrows for intermolecular noise.}
	\label{fig:ising}
\end{figure}

Firstly, we performed atomistic molecular dynamics simulations of four commodity polymer melts: polyethylene (PE), isotactic polypropylene (PP), atactic poly(methyl methacrylate) (PMMA), and atactic polystyrene (PS). The topological dissipation mechanism is quantified via intrachain momentum correlations by calculating the velocity cross-correlation function,
\begin{equation}
	C_\text{intra}(t;\Delta n) = \langle \mathbf{v}_i(0) \cdot \mathbf{v}_{j}(t) \rangle,
\label{eq:vccf}
\end{equation}
where $\mathbf{v}_i$ denotes the velocity of monomer $i$ (or dimer for PE), and $\Delta n = |i-j|$ is the backbone separation. We found that the peak VCCF values, plotted in Fig.~\ref{fig:ising}(a), decay exponentially with $\Delta n$ that is consistent with the intramolecular Ising-like form of dynamical acceleration mechanism induced by nanoparticle-segment interaction in our previous study of polymer nanocomposite~\cite{chentaoNC2019}. Remarkably, the ratio of Kuhn length $n_\text{k}$~\cite{everaers2020kremer} to dynamical correlation length $n_\text{d}$ converges to $\sim$3.0 for all four polymers, as shown in the inset of Fig.~\ref{fig:ising}(a), suggesting a universal scaling between chain stiffness and dynamical coupling in studied polymers. 
\par
On the other hand, intramolecular correlations are found much stronger than intermolecular counterpart, as shown in Supplemental Material (SM) Fig.~S1. Therefore we assume that topological dissipation and conventional intermolecular spatial friction are uncorrelated, 
\begin{eqnarray}
	\langle (\delta_i^{\tau} + \delta_i^{\xi}) (\delta_j^{\tau} + \delta_j^{\xi})^T \rangle = 
	\langle \delta_i^{\xi} (\delta_j^{\xi})^T \rangle + \langle \delta_i^{\tau} (\delta_j^{\tau})^T \rangle
\label{eq:decompose}
\end{eqnarray}
here $\delta_i^{\tau}$ and $\delta_i^{\xi}$ represent topological and conventional Langevin-type spatial noises respectively. To quantify the relatively magnitude of each components, we define the fraction $\Lambda = \zeta^{\tau}/(\zeta^\tau + \zeta^\xi)$, where $\zeta^\tau$ and $\zeta^\xi$ are the corresponding friction coefficients contributed by two dissipative components respectively. While direct measurement of $\Lambda$ remains challenging, we introduce a computationally accessible proxy $\lambda = \mathcal{I}^\tau/(\mathcal{I}^\xi + \mathcal{I}^\tau)$. This metric combines two physically distinct contributions: (i) the intermolecular correlation integral $\mathcal{I}^\xi = \rho \int_0^\infty 4\pi r^2 g(r) \max_t C_\text{inter}(t;r) \, dr$, which captures spatial momentum transfer between chains through the pair correlation function $g(r)$ and inter-chain velocity cross-correlation function $C_\text{inter}(t;r)$, and (ii) the intramolecular sum $\mathcal{I}^\tau = \sum_{\Delta n=1}^{N-1} \max_t [C_\text{intra}(t;\Delta n)] \cdot [2(N-\Delta n)/N]$, quantifying directional momentum propagation along chains via the intra-chain correlation function $C_\text{intra}(t;\Delta n)$. For the PP melt ($N = 40$) we studied, atomistic simulations yield $\lambda \approx 0.50$, demonstrating comparable topological and spatial friction—a finding consistent with Ref.~\onlinecite{lyubimov2011first}. The $\lambda$ value decreases systematically with coarse-graining level (Table~\ref{table1}) but shows a weak chain-length dependence (Table S2), confirming that topological dissipation is significant at moderate CG levels. According to the ratio $n_\text{d}/n_\text{k}$, the $\lambda$ is expected to converge to zero at the Kuhn CG level for high-temperature melts, which implies a reduction to Rouse-like dynamics.

\begin{table}[b]
	\caption{\label{table1} The $\lambda$ value for different polymer melt with the degree of polymerization 40. The ratio in the parentheses represent the coarse-grainning level, i.e., number of monomers per CG bead. Notably, in the PE system, a dimer (two monomers) is mapped to one CG bead, whereas for all other polymers, each CG bead represents a single monomer.}
	\begin{ruledtabular}
		\begin{tabular}{cccc}
			PP(1:1) & PE(2:1) & PMMA(1:1) & PS(1:1) \\
			\colrule
			0.497 & 0.386 & 0.287 & 0.279 \\
		\end{tabular}
	\end{ruledtabular}
\end{table}
\par
According to Eq.~\ref{eq:decompose}, we design a Markovian model as follows,
\begin{eqnarray}
	\mathbf{F}_{i} &=& \mathbf{F}_{i}^{\text{C}} +  \mathbf{F}_{i}^{\xi} + \mathbf{F}_{i}^{\tau} \nonumber \\
	&=& \mathbf{F}_{i}^{\text{C}} +  \mathbf{F}_{i}^{\xi} + \sum_{j \in \mathcal{T}_i} (\mathbf{F}_{ij}^{\mathcal{D}} + \mathbf{F}_{ij}^{\mathcal{R}})
	\label{eq:EOM}
\end{eqnarray}
where $\mathbf{F}_i^\text{C}$ is the conservative force between CG beads. $\mathbf{F}_i^\xi$ represents intermolecular non-conservative forces, including frictions and random forces. While $\mathbf{F}_{i}^{\tau}$ is for the summation of intramolecular dissipative forces ($\mathbf{F}_{ij}^{\mathcal{D}}$) and random  forces ($\mathbf{F}_{ij}^{\mathcal{R}}$),  among topologically connected bead sets $\mathcal{T}_i$. Such a decomposition scheme is schematically illustrated in Fig.~\ref{fig:ising}(b), where the orange and black arrows distinctly represent the intrachain and interchain dissipation channels, respectively. 
\par
Based on the results in Fig.~\ref{fig:ising}(a), we construct the topological random forces using an Ising-type ansatz with a pairwise form,
\begin{equation}
\langle [\mathbf{F}_{ij}^\mathcal{R}(t)][\mathbf{F}_{ij}^\mathcal{R}(0)]^T\rangle = 2A e^{-|i-j|/n_\text{d}}\delta(t)/\beta,
\label{eq:random}
\end{equation}
where $n_\text{d}$ is the dynamical correlation length, and $\beta = 1/k_\text{B} T$. Numerically, we generate the random force by $\mathbf{F}_{ij}^{\mathcal{R}} = \sqrt{\frac{2 A \exp{(- \lvert i-j \rvert/n_\text{d})}}{\beta \Delta t}} (\Theta_{ij} \mathbf{I} + \sqrt{2}\mathbf{W}_{ij}^{A})\cdot \hat{\mathbf{r}}_{ij}$, $\Delta t$ is the integrated timestep of the simulation, $\mathbf{r}_{ij} = \mathbf{r}_{i} - \mathbf{r}_{j}$ and $\hat{\mathbf{r}}_{ij} = \mathbf{r}_{ij} / \lvert \mathbf{r}_{ij} \rvert$; $\Theta_{ij}(t)$ is a one-dimensional Gaussian white noise, and the $\mathbf{W}_{ij}^{A}=\frac{1}{2}(\mathbf{W}_{ij}^{\mu \nu}-\mathbf{W}_{ij}^{\nu \mu})$ is an anti-symmetric $3 \times 3$ noise matrix constructed from a set of independent Wiener increments $\mathbf{W}_{ij}$ \cite{li2015incorporation}. To satisfy the fluctuation-dissipation theorem, the dissipative force takes the following form with $\mathbf{v}_{ij} = \mathbf{v}_{i} - \mathbf{v}_{j}$,
\begin{equation}
\mathbf{F}_{ij}^\mathcal{D} = -A e^{-|i-j|/n_\text{d}}\mathbf{v}_{ij}.
\label{eq:dissipative}
\end{equation}
For intermolecular non-conservative forces $\mathbf{F}_i^\xi$, one can implement the popular forms, such as standard Langevin dynamics~\cite{grest1986molecular} with $\mathbf{F}_i^\xi = -\gamma\mathbf{v}_i + \mathbf{F}_i^\text{R}$, or dissipative particle dynamics~\cite{groot1997dissipative} where $\mathbf{F}_i^\xi = \sum_{j\in\mathcal{S}_i} -\gamma w^\text{D}(r_{ij})(\hat{\mathbf{r}}_{ij}\cdot\mathbf{v}_{ij})\hat{\mathbf{r}}_{ij} + \mathbf{F}_i^\text{R}$.
\par
Note that Eqs.~\ref{eq:EOM},~\ref{eq:random}, and \ref{eq:dissipative} are only applicable to macromolecular systems with chain-like topology. For non-bonded liquids, star-polymers for instance, which were widely investigated in literature ~\cite{li2015incorporation, li2017computing, klippenstein2021cross, huan2023construction}, memory kernels should be explicitly adopted for constructing non-Markovian dynamics. As a prototype system, we simulated here a unentangled PP melt with polymerization degree $N=40$, where each propylene monomer is coarse-grained into a single bead. The conservative forces $\mathbf{F}_i^\text{C}$ between beads are derived via iterative Boltzmann inversion~\cite{IBI2003}, while non-conservative forces are parameterized following the philosophy outlined in the above section (see SM for details). To check the performance of the model, we performed both Langevin (LE) and dissipative particle dynamics (DPD) simulations. To distinguish them from conventional LE or DPD formalism, they are labelled as $i$-LE and $i$-DPD, the prefix $i$- denotes both intramolecular dissipation and Ising-type exponential decay. For benchmark purposes, conventional LE or DPD simulations without topological dissipation but with identical total friction coefficient are performed.
\begin{figure}
	\includegraphics[width=\linewidth]{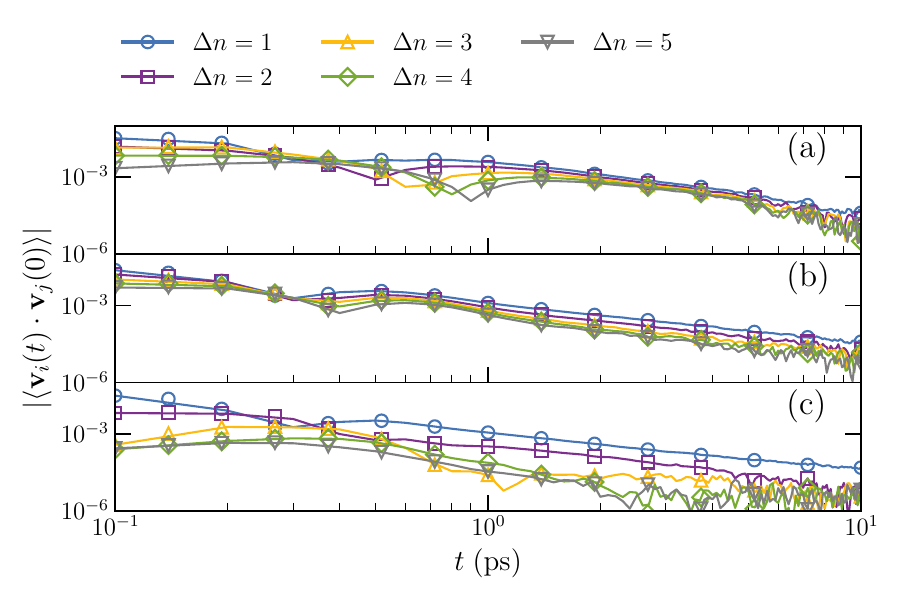}
	\caption{\label{VCCF} The $ \vert C_\text{intra}(t;\Delta n) \rvert$ calculated from (a) atomistic simulations, (b) $i$-LE simulations, and (c) LE simulations. The parameters used for $i$-LE simulations are $\gamma =  169.2 \; \text{amu} \cdot \text{ps}^{-1}$ and $A = 343.5 \; \text{amu} \cdot \text{ps}^{-1}$, while the parameter for the LE simulations is $\gamma = 337.0 \; \text{amu} \cdot \text{ps}^{-1}$.}
\end{figure}
\begin{figure}
	\includegraphics[width=\linewidth]{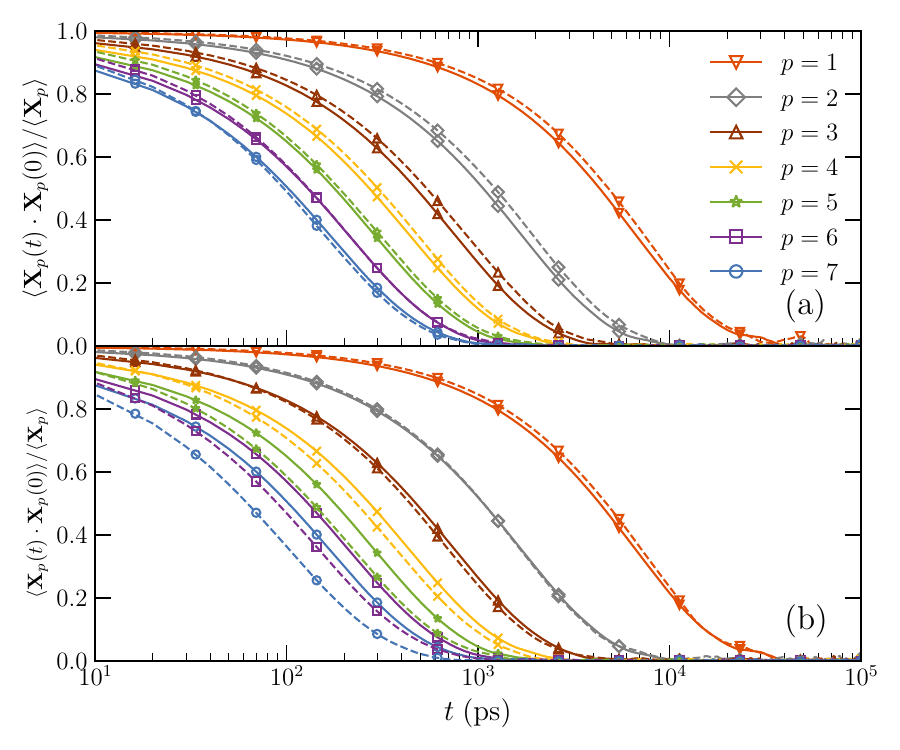}
	\caption{\label{RouseMode} The autocorrelation function of Rouse modes $\langle \mathbf{X}_p (t) \cdot \mathbf{X}_p(0) \rangle / \langle \mathbf{X}_p \rangle$ for $p=1 \text{ to } 7$. (a) atomistic simulations (solid line) vs. $i$-LE simulations (dashed lines). (b) atomistic simulations (solid line) vs. LE simulations (dashed lines).}
\end{figure}
\par
To validate the preservation of topological correlations, we performed quantitative comparisons of the intrachain velocity cross-correlation function $C_\text{intra}(t;\Delta n)$ across different simulation methods. Figures~\ref{VCCF} ($i$-LE vs.\ LE) and S7 ($i$-DPD vs.\ DPD) reveal persistent correlations in atomistic simulations even at substantial monomer separations ($\Delta n = 5$). Our $i$-LE and $i$-DPD models successfully reproduce this behavior, whereas conventional LE and DPD simulations exhibit rapidly decaying correlations with increasing $\Delta n$, as topological correlations are maintained solely through conservative interactions (bond, angle, and torsion potentials).
\par
We further quantified topological preservation through Rouse mode analysis:
\begin{equation}
\mathbf{X}_p(t) \equiv \frac{1}{N}\sum_{i=1}^N \cos\left[\frac{p\pi}{N}\left(i-\frac{1}{2}\right)\right]\mathbf{r}_i(t)
\end{equation}
for $N=40$ PP chains. Figure~\ref{RouseMode}(a) demonstrates excellent agreement for modes $p=1$--$7$ in $i$-LE simulations, while Fig.~\ref{RouseMode}(b) shows significant deviations in higher modes ($p\geq4$) for LE simulations due to missing topological dissipation. The agreement in lower modes arises since such modes involve chain segments much larger than $n_\text{d}$ where topological dynamical correlation along chain backbone vanishes. Similar results are observed for $i$-DPD vs.\ DPD comparisons (Fig.~S8).
\par

\begin{figure}[b]
	\includegraphics[width=\linewidth]{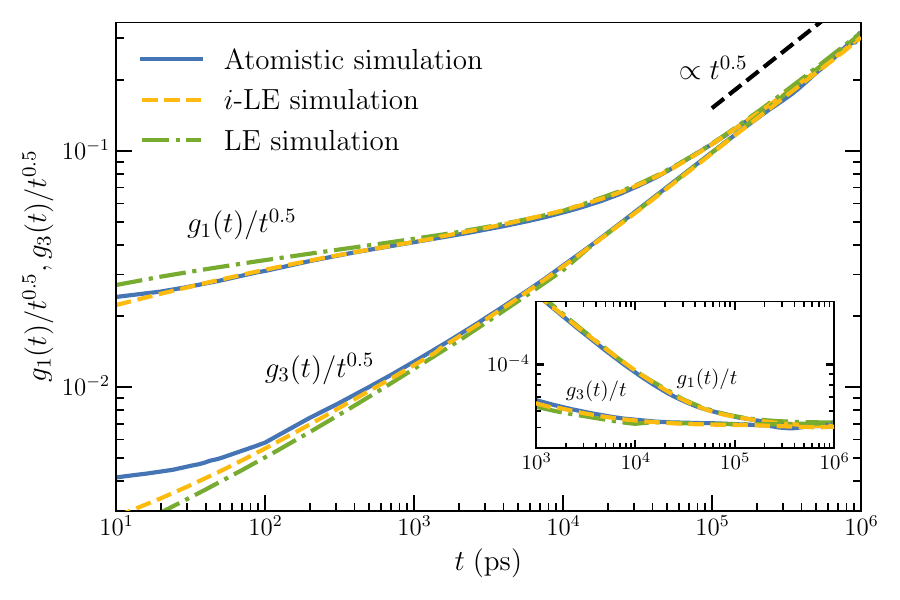}
	\caption{\label{MSD}The $g_1(t)$ and $g_3(t)$ divided by $t^{0.5}$ calculated from atomistic, $i$-LE and LE simulations. The inset shows the results of the $g_1(t)$ and $g_3(t)$ divided by $t$.}
\end{figure}
\par
To further validate dynamical properties, we calculated the mean square displacement (MSD) for both monomers $g_1(t)$ and polymer center-of-mass $g_3(t)$, as shown in Figure~\ref{MSD}.  The $i$-LE simulations correct the subdiffusive behavior within 1 ns, where significant deviations are observed between LE simulations and the atomistic reference model. These results are consistent with Rouse mode analysis in Fig.~\ref{RouseMode}. Note that deviations emerge in the short-time regime ($t < 20$ ps) in $g_1(t)$ due to unresolved high-frequency atomic-scale vibrations, requiring generalized Langevin dynamics with memory kernels for accurate description~\cite{zhang2024chemically}. Nevertheless, we still observed improve consistency in monomer velocity auto-correlation functions after the introduction of topological dissipation as shown in Figure~S6, although sub-picosecond dynamics still demand memory kernel treatments. Additionally, the deviation in $g_3(t)$ before $\sim$100 ps arises from the lack of hydrodynamic interactions (HIs) in the LE simulations \cite{farago2011anomalous}. The $i$-DPD model naturally accounts for HIs, demonstrating superior agreement in short-time diffusion behavior (Fig.~S9). The current model demonstrates more consistent diffusion behavior compared to earlier Mori-Zwanzig-based models that lacked the topological dependence of friction \cite{lemarchand2017coarse, deichmann2018bottom, zhang2024chemically}.

\par
\begin{figure}
	\includegraphics[width=\linewidth]{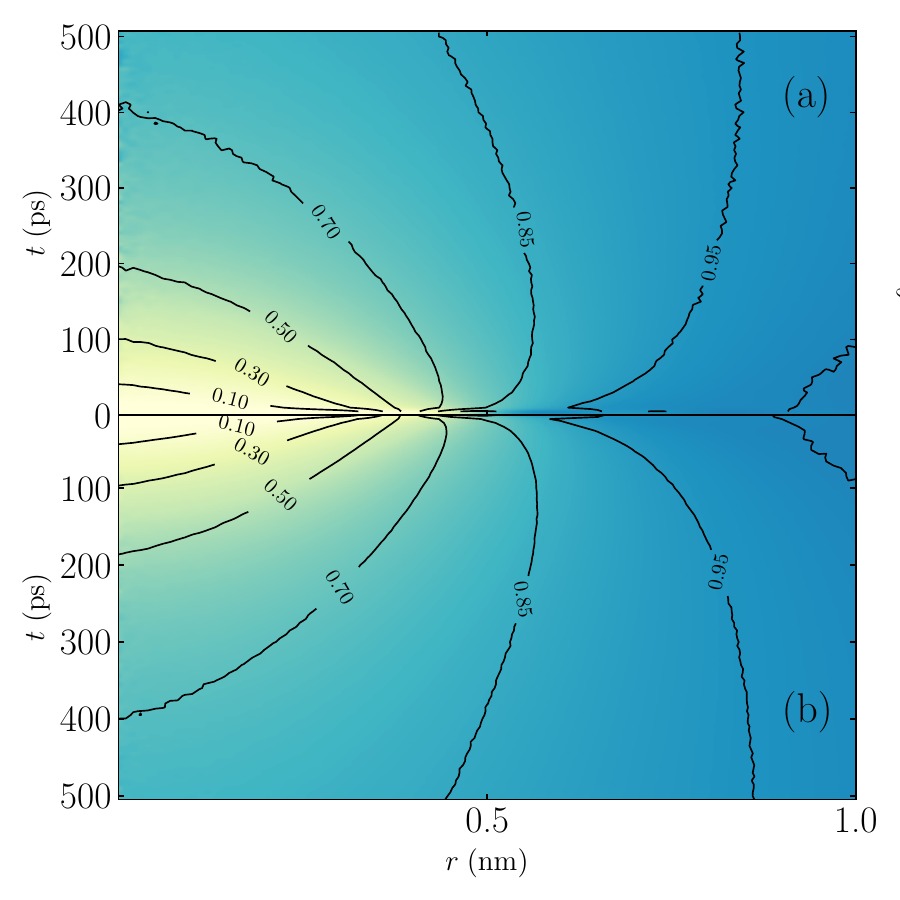}
	\caption{\label{vanhove}The distinct part of van Hove function calculated from (a) atomistic and (b) $i$-LE simulations.}
\end{figure}

Beyond topology-dependent dynamics, we systematically validated multiscale collective dynamics. The distinct van Hove function, $G_d(r,t) = \frac{1}{4\pi\rho N_p r^2} \left\langle \sum_{i=1}^{N_p} \sum_{j\neq i}^{N_p} \delta(r-|\mathbf{r}_i(t)-\mathbf{r}_j(0)|) \right\rangle$, quantifies interparticle spatial correlations (Fig.~\ref{vanhove}; Fig.~S11 for $i$-DPD/DPD comparison). At $t=0$, $G_d(r,t)$ reduces to the radial distribution function governed by conservative interactions. The $i$-LE simulations accurately reproduce both structural correlations and their temporal evolution observed in atomistic simulations. The self-part, $G_s(r,t) = \frac{1}{N_p} \left\langle \sum_{i=1}^{N_p} \delta(r-|\mathbf{r}_i(t)-\mathbf{r}_i(0)|) \right\rangle$, and its Fourier transform, the incoherent intermediate scattering function, $F_s(q,t) = \frac{1}{N_p} \left\langle \sum_{i=1}^{N_p} e^{-i\mathbf{q}\cdot[\mathbf{r}_i(t)-\mathbf{r}_i(0)]} \right\rangle$, exhibit consistent dynamics from ballistic to diffusive regimes (Figs.~S12, S13). At larger scales, the stress relaxation modulus, $G(t) = \frac{V}{k_BT} \langle \sigma_{\alpha\beta}(t)\sigma_{\alpha\beta}(0) \rangle$ and zero-shear viscosity show good agreement (Fig.~S14, Table~S4). Because the current model naturally captures chain-end effects, we also extended the chain length into the entangled regime and compute diffusion coefficients that closely match experimental values (Fig.~S15)~\cite{von2009effect}. This demonstrates our model's bottom-up universality---even though its parameters were developed from unentangled melts.
\par

In summary, we have identified topological dissipation as an essential mechanism governing multiscale dynamics in coarse-grained polymer systems. Our analysis of various atomistic polymer melt simulations reveals a universal Ising-type exponential decay in topological correlations. Building on this discovery, we establish an efficient coarse-grained molecular dynamics method based on the Langevin or DPD formalism that eliminates the need for both complex memory kernel reconstruction and computationally demanding convolution operations. Extensive validation demonstrates that our method successfully reproduces a wide range of collective dynamical properties across multiple spatiotemporal scales. These findings advance both the theoretical understanding of polymer dynamics and provide practical methodological innovations for complex fluid simulations. The principles established here may inspire new directions in coarse-grained modeling of other topologically complex systems.

\begin{acknowledgments}
\textit{Acknowledgments}---This research is supported by the National Key Research and Development Program of China (2023YFB3812801) and the National Natural Science Foundation of China (22473049, and 22133002). H.-J. Q. and Z.-Y. L. also acknowledge the support from the Program for JLU Science and Technology Innovative Research Team.
\end{acknowledgments}
\bibliography{ref}

\end{document}